# Best Practices for Applying Deep Learning to Novel Applications

**Leslie N. Smith**

Navy Center for Applied Research in Artificial Intelligence

U.S. Naval Research Laboratory, Code 5514

Washington, DC 20375

leslie.smith@nrl.navy.mil

**ABSTRACT**

This report is targeted to groups who are subject matter experts in their application but deep learning novices. It contains practical advice for those interested in testing the use of deep neural networks on applications that are novel for deep learning. We suggest making your project more manageable by dividing it into phases. For each phase this report contains numerous recommendations and insights to assist novice practitioners.

**Introduction**

Although my focus is on deep learning (DL) research, I am finding that more and more frequently I am being asked to help groups without much DL experience who want to try deep learning on their novel (for DL) application. The motivation for this NRL report derives from noticing that much of my advice and guidance is similar for all such groups. Hence, this report discusses the aspects of applying DL that are more universally relevant.

While there are several useful sources of advice on best practices for machine learning [1-5], there are differences relevant to DL that this report addresses. Still, I recommend the reader read and become familiar with these references as they contain numerous gems. In addition, there are many sources on best practices on the topic of software engineering and agile methodologies that I assume the reader is already familiar with (e.g., [6, 7]). The closest reference to the material in this report can be found in Chapter 11 of "Deep Learning" [8] on "Practical Methodology" but here I discuss a number of factors and insights not covered in this textbook.

You can see below that a deep learning application project is divided into phases. However, in practice you are likely to find it helpful to return to an earlier phase. For example, while finding an analogy in phase 3, you might discover new metrics that you hadn't considered in phase 1. All of these best practices implicitly include iteratively returning to a phase and continuous improvement as the project proceeds.

**Phase 1: Getting prepared**

In this report I assume you are (or have access to) a subject matter expert for your application. You should be familiar with the literature and research for solving the associated problem and know the state-of-the-art solutions and performance levels. I recommend you consider here at the beginning if a deep learning solution is a worthwhile effort. You must consider the performance level of the state-of-the-art and if it is high, whether it is worthwhile to put in the efforts outlined in this report for an incremental improvement. Don't jump into deep learning only because it seems like the latest and greatest methodology.  You should also consider if you have the computer resources since each job to train a deep network will likely take days or weeks.  I have made ample use of DoD's HPC systems in my own research.    In addition, you should consider if machine learning is appropriate at all – remember training a deep network requires lots of labeled data, as described in phase 2.

The first step is quantitatively defining what success looks like.  What will you see if this is successful, whether it is done by human or machine? This helps define your evaluation metrics.  Which metrics are important?  Which are less important?  You need to specify all quantitative values that play a role in the success of this project and determine how to weigh each of them.  You also need to define objectives for your metrics; is your goal surpass human level performance?  Your objectives will strongly influence the course of the project.  Knowing quantitatively what human performance is on this task should guide your objectives; how does the state-of-the-art compare to human performance?  Also, knowing how a human solves this task will provide valuable information on how the machine might solve it.

Some of these metrics can also lead to the design of the loss function, which is instrumental in guiding the training of the networks.  Don't feel obligated to only use softmax/cross entropy/log loss just because that is the most common loss function, although you should probably start with it.  Your evaluation metrics are by definition the quantities that are important for your application.  Be willing to test these metrics as weighted components of the loss function to guide the training (see phase 6).

Although you are likely considering deep learning because of its power, consider how to **make the network's "job" as easy as possible**.  This is anti-intuitive because it is the power of deep networks that likely motivates you to try it out.  However, the easier the job that the networks must perform, the easier it will be to train and the better the performance.    Are you (or the state-of-the-art) currently using heuristics/physics that can be utilized here?  Can the data be preprocessed?  While the network can learn complex relationships, remember: "the easier the network's job, the better it will perform".  So it is worthwhile to spend time considering what you can leverage from previous work and what the network needs to do for you.  Let's say you want to improve on a complex process where the physics is highly approximated (i.e., a "spherical cow" situation); you have a choice to input the data into a deep network that will (hopefully) output the desired result or you can train the network to find the correction in the approximate result.   The latter method will almost certainly outperform the former.  On the other hand, do not rely on manual effort to define potential heuristics – the scarcest resource is human time so let the network learn its representations rather than require any fixed, manual preprocessing.

In addition, you might want to write down any assumptions or expectations you have regarding this state-of-the-art process as it will clarify them for yourself.

**Phase 2: Preparing your data**

Deep learning requires a great deal of training data.  You are probably wondering "how much training data do I need?"   The number of parameters in the network is correlated with the amount of training data. The number of training samples will limit your architectural choices in phase 6.  The more training data, the larger and more accurate the network can be.  So the amount of training data depends on the objectives you defined in phase 1.

In addition to training data, you will need a smaller amount of labeled validation or test data. This test data should be similar to the training data but not the same. The network is not trained on the test data but it is used to test the generalization ability of the network.

If the amount of training data is very limited, consider transfer learning [9] and domain adaptation [10, 11]. If this is appropriate, download datasets that are closest to your data to use for pre-training. In addition, consider creating synthetic data. Synthetic data has the advantages that you can create plenty of samples and make it diverse.

The project objectives also guides the choosing of the training data samples. Be certain that the training data is directly relevant to the task and that it is diverse enough that it **covers the problem space**. Study the statistics of each class. For example, are the classes balanced? An example of a balanced class is cats versus dogs while an unbalanced class with be cats versus all other mammals (if your problem is inherently unbalanced, talk with a deep learning expert).

What preprocessing is possible? Can you zero mean and normalize the data? This makes the network's job easier as it removes the job of learning the mean. Normalization also makes the network's job easier by creating greater similarity between training samples.

As discussed above, investigate if there are ways to lower the dimensionality of the data using a priori knowledge or known heuristics. You don't need to spend time to manually determine heuristics because the goal is to save human time and you can let the network learn its own representations. Just know that the more irrelevant data the network has to sift through, the more training data is needed and the more time it will take to train the network. So leverage what you can from prior art.

**Phase 3: Find an analogy between your application and the closest deep learning applications**

Experts know not to start from scratch for every project. This is what makes them experts. They reuse solutions that have worked in the past and they search the deep learning literature for solutions from other researchers. Even if no one has ever done what you are trying to do, you still need to leverage whatever you can from the experts.

Deep learning has been applied to a variety of applications. In order to create your baseline model – your starting point – you need to find the applications that are in some ways similar to your application. You should search the DL literature and consider the "problems" various applications are solving to compare with the "problem" you need to solve in your application. Find similarities and analogies between these problems. Also, note of the differences between your new application and the deep learning application because these differences might require changing the architecture in phase 6.

When you find the closest application, look for code to download. Many researchers make their code available when they publish in a wide spread effort to release reproducible research. Your first aim is to replicate the results in the paper of the closest application. Later you should modify various aspects to see the effects on the results in a "getting to know it" stage. If you are lucky, there will be several codes available and you should replicate the results for all of them. This comparison will provide you with enough information so you can create a baseline in phase 4.

There are a few "classic" applications of deep learning and well known solutions. These include image classification/object recognition (convolutional networks), processing sequential data (RNN/LSTM/GRU)

Version 1.0                                                                                                                              2/27/17

such as language processing, and complex decision making (deep reinforcement learning). There are also a number of other applications that are common, such as image segmentation and super-resolution (fully convolutional networks) and similarity matching (Siamese networks). Appendix A lists a number of recent deep learning applications, the architecture used, and links to the papers that describe this application. This can give you some ideas but should not be your source for finding deep learning applications. Instead you should carefully search https://scholar.google.com and https://arxiv.org for the deep learning applications.

**Phase 4: Create a simple baseline model**

**Always start simple, small, and easy**. Use a smaller architecture than you anticipate you might need. Start with a common objective function. Use common settings for the hyper-parameters. Use only part of the training data. This is a good place to adopt some of the practices of agile software methodologies, such as simple design, unit testing, and short releases. Only get the basic functionality now and improve on it during phase 6. That is, plan on small steps, continuous updates, and to iterate.

Choose only one of the common frameworks, such as Caffe, TensorFlow, or MXnet. Plan to only use one framework and one computer language to minimize errors from unnecessary complexity. The framework and language choice will likely be driven by the replication effort you performed in phase 3.

If the network will be part of a larger framework, here is a good place to check that the framework APIs are working properly.

**Phase 5: Create visualization and debugging tools**

Understanding what is happening in your model will affect the success of your project. Carpenters have an expression "measure twice, cut once". You should think "code once, measure twice". In addition to evaluating the output, you should visualize your architecture and measure internal entities to understand why you are getting the results you are obtaining. Without diagnostics, you will be shooting in the dark to fix problems or improve performance.

You should have a general understanding of problems related to high bias (converging on the wrong result) versus high variance (not converging well) because there are different solutions for each type of problem; for example, you might fix high bias problems with a larger network but you would handle high variance problems by increasing the size of your training dataset.

Set up visualizations so you can monitor as much as possible while the architecture evolves. When possible, set up unit tests for all of your code modifications. You should compare training error to test error and both to human level performance. You might find your network will behave strangely and you need ways to determine what is going on and why. Start debugging the worst problems first. Find out if the problems are with the training data, aspects of the architecture, or the loss function.

Keep in mind that error analysis tries to explain the difference between current performance and perfect performance. Ablative analysis tries to explain the difference between some baseline performance and current performance. One or the other or both can be useful.

One motivation for using TensorFlow as your framework is that it has a visualization system called TensorBoard that is part of the framework. One can output the necessary files from TensorFlow and TensorBoard can be used to visualize your architecture, monitor the weights and feature maps, and explore the embedded space the network creates. Hence, the debugging and visualization tools are available in the framework. With other frameworks, you need to find these tools (they are often available online) or create your own.

**Phase 6: Fine tune your model**

This phase will likely take the most time. You should experiment extensively. And not just with factors you believe will improve the result but try changing every factor just to learn what happens when it changes. Change the architecture design, depth, width, pathways, weight initialization, loss function, etc. Change each hyper-parameter to learn what the effect of increasing or decreasing it is. I recommend using the learning rate range test [12] to learn about the behavior of your network over a large range of learning rates. A similar program can be made to study the effect of other hyper-parameters.

Try various regularization methods, such as data augmentation, dropout, and weight decay. Generalization is one of the key advantages of deep networks so be certain to test regularization methods in order to maximize this ability to generalize to unseen cases.

You should experiment with the loss function. You used a simple loss function in the baseline but you also created several evaluation metrics that you care about and define success. **The only difference between the evaluation metrics and the loss function is that the metrics apply to the test data and the loss function is applied to the training data in order to train the network.** Can a more complicated loss function produce a more successful result? You can add weighted components to the loss function to reflect the importance of each metric to the results. Just be very careful to not complicate the loss function with unimportant criterion because it is the heart of your model.

Earlier you found analogies between your application and existing deep learning applications and chose the closest to be your baseline. Now compare to the second closest application. Or the third. What happens if you follow another analogy and use that architecture? Can you imagine a combination of the two to test?

In the beginning, you should have successes from some low hanging fruit. As you go on it will become more difficult to improve the performance. The objectives you defined in phase 1 should guide how far you want to pursue the performance improvements. Or you might want to now revise the objectives your defined earlier.

**Phase 7: End-to-end training, ensembles and other complexities**

If you have the time and budget, you can investigate more complex methods and there are worlds of complexities that are possible. There exists a huge amount of deep learning literature and more papers are appearing daily. Most of these papers declare new state-of-the-art results with one twist or another and some might provide you a performance boost. This section alone could fill a long report because there are so many architectures and other options to consider but if you are at this stage, consider talking with someone with a great deal of deep learning expertise because advice at this stage is likely to be unique to your application. .

However, there are two common methods you might consider; end-to-end training and ensembles.

As a general rule, end-to-end training of a connected system will outperform a system with multiple parts because a combined system with end-to-end training allows each of the parts to adapt to the task. Hence, it is useful to consider combining parts, if it is relevant for your application.

Ensembles of diverse learners (i.e., bagging, boosting, stacking) can also improve the performance over a single model. However, this will require you to train and maintain all the members of the ensemble. If your performance objectives warrant this, it is worthwhile to test an ensemble approach.

**Summary**

This report lays out many factors for you to consider when experimenting with deep learning on an application where it hasn't been previously used. Not every item here will be relevant but I hope that it covers most of the factors you should consider during a project. I wish you much luck and success in your efforts.

**Appendix A:  Table of various deep learning applications**

The following table lists some recent applications of deep learning, the architecture used for this application and a few references to papers in the literature that describe the application in much more detail.

| Application | Architecture | Comments |
| --- | --- | --- |
| Colorization of Black and White Images. | Large, fully convolutional | http://www.cs.cityu.edu.hk/~qiyang/publications/iccv-15.pdf <br><br> http://arxiv.org/pdf/1603.08511.pdf |
| Adding Sounds To Silent Movies. | CNN + LSTM | http://arxiv.org/pdf/1512.08512.pdf |
| Automatic Machine Translation. | Stacked networks of large LSTM recurrent neural networks | http://www.nlpr.ia.ac.cn/cip/ZongPublications/2015/IEEE-Zhang-8-5.pdf <br> https://arxiv.org/abs/1612.06897 <br> https://arxiv.org/abs/1611.04558 |
| Object Classification in Photographs. | Residual CNNs, ResNeXt, Densenets | http://papers.nips.cc/paper/5207-deep-neural-networks-for-object-detection.pdf |
| Automatic Handwriting Generation. | RNNs | http://arxiv.org/pdf/1308.0850v5.pdf |
| Character Text Generation. | RNNs | http://arxiv.org/pdf/1308.0850v5.pdf |
| Image Caption Generation. | CNN + LSTM | http://arxiv.org/pdf/1505.00487v3.pdf |
| Automatic Game Playing. | Reinforcement learning + CNNs | http://www.nature.com/nature/journal/v529/n7587/full/nature16961.html <br> https://arxiv.org/abs/1612.00380 |
| Generating audio | WaveNet = Dilated PixelCNN | https://arxiv.org/abs/1609.03499 <br> https://arxiv.org/abs/1612.07837 <br> https://arxiv.org/abs/1610.09001 |
| Object tracking | CNN + hierarchical LSTMs | https://arxiv.org/abs/1701.01909 <br> https://arxiv.org/abs/1611.06878 <br> https://arxiv.org/abs/1611.05666 |
| Lip reading | CNN + LSTMs | https://arxiv.org/abs/1701.05847 <br> https://arxiv.org/abs/1611.05358 |
| Modifying synthetic data into labeled training data | GAN | https://arxiv.org/abs/1701.05524 |
| Single image super-resolution | Deep, fully convolutional networks | https://arxiv.org/abs/1612.07919 |
| | | https://arxiv.org/abs/1611.03679 |
| | | https://arxiv.org/abs/1511.04587 |
| | | https://arxiv.org/abs/1611.00591 |
| Speech recognition | LSTMs | https://arxiv.org/abs/1701.03360 |
| | | https://arxiv.org/abs/1701.02720 |
| Generate molecular structures | RNNs | https://arxiv.org/abs/1701.01329 |
| Time series analysis | Resnet + RNNs | https://arxiv.org/abs/1701.01887 |
| | | https://arxiv.org/abs/1611.06455 |
| Intrusion detection | RNNs or CNNs | https://arxiv.org/abs/1701.02145 |
| Autonomous Planning | Predictron architecture | https://arxiv.org/abs/1612.08810 |
| Object detection | | https://arxiv.org/abs/1612.08242 |
| Multi-modality classification | CNN + GAN | https://arxiv.org/abs/1612.07976 |
| | | https://arxiv.org/abs/1612.00377 |
| | | https://arxiv.org/abs/1611.06306 |
| Health monitoring | All are used | https://arxiv.org/abs/1612.07640 |
| Robotics | CNN (perception) RL (control) | https://arxiv.org/abs/1612.07139 |
| | | https://arxiv.org/abs/1611.00201 |
| Domain adaptation | | https://arxiv.org/abs/1612.06897 |
| Self-driving | CNNs | https://arxiv.org/abs/1612.06573 |
| | | https://arxiv.org/abs/1611.08788 |
| | | https://arxiv.org/abs/1611.05418 |
| Visual question answering | Resnet | https://arxiv.org/abs/1612.05386 |

|  |  | https://arxiv.org/abs/1611.01604 |
|  |  | https://arxiv.org/abs/1611.05896 |
|  |  | https://arxiv.org/abs/1611.05546 |
| Weather prediction | Graphical RNN | https://arxiv.org/abs/1612.05054 |
| Detecting cancer | RBM | https://arxiv.org/abs/1612.03211 |
| Genomics | Multiple NNs | https://arxiv.org/abs/1611.09340 |
| Semantic segmentation | Fully conv Densenet | https://arxiv.org/abs/1611.09326 |
|  |  | https://arxiv.org/abs/1611.06612 |
| Hyperspectral classification | CNN | https://arxiv.org/abs/1611.09007 |
| Natural Language Processing (NLP) | LSTM, GRU | https://arxiv.org/abs/1606.0673 |
| Face detection | CNN | https://arxiv.org/abs/1611.00851 |